# Spectroscopy and level detuning of few-electron spin states in parallel InAs quantum dots

Claes Thelander,[1,*] Malin Nilsson,[1] Florinda Viñas Boström,[1] Adam Burke,[1] Sebastian Lehmann,[1] Kimberly A. Dick,[1,2] and Martin Leijnse[1]

[1]*Division of Solid State Physics and NanoLund, Lund University, Box 118, S-221 00 Lund, Sweden*

[2]*Center for Analysis and Synthesis, Lund University, Box 124, S-221 00 Lund, Sweden*

We use tunneling spectroscopy to study the evolution of few-electron spin states in parallel InAs nanowire double quantum dots (QDs) as a function of level detuning and applied magnetic field. Compared to the much more studied serial configuration, parallel coupling of the QDs to source and drain greatly expands the probing range of excited state transport. Owing to a strong confinement, we can here isolate transport involving only the very first interacting single QD orbital pair. For the (2,0) – (1,1) charge transition, with relevance for spin-based qubits, we investigate the excited (1,1) triplet, and hybridization of the (2,0) and (1,1) singlets. An applied magnetic field splits the (1,1) triplet, and due to spin-orbit induced mixing with the (2,0) singlet, we clearly resolve transport through all triplet states near the avoided singlet-triplet crossings. Transport calculations, based on a simple model with one orbital on each QD, fully replicate the experimental data. Finally, we observe an expected mirrored symmetry between the 1–2 and 2–3 electron transitions resulting from the two-fold spin degeneracy of the orbitals.

I.  INTRODUCTION

Tunneling spectroscopy applied to quantum dot (QD) systems is today a standard tool to extract information on electron orbitals and spins, and how these relate to various material properties [1-3]. In recent years, there has been a particular interest in tunneling spectroscopy of double QDs (DQDs) in materials with strong spin-orbit (SO) interaction [4-6]. Such interaction provides large, orbital-dependent |*g*|-factors in QDs [3-7], important to various qubit concepts building on manipulation of individual spins [8] or on Majorana states [9]. In most studies involving DQDs, the QDs are oriented serially, one after the other, relative to a source and drain contact. One reason for this focus was that many new materials were first synthesized into narrow, elongated objects, such as nanowires or nanotubes [4-6, 10]. Another motivation, important for devices, is that the serial DQD configuration enables probing of spin-states through Pauli spin blockade [11-12].

At zero bias, transport in serial DQDs only occurs when states in both QDs, tuned by local gates, align with the contact chemical potential at so-called triple degeneracy points. With increasing source-drain bias, these points evolve into triangular windows, where sequential tunneling through excited states is also possible. However, from the point-of-view of tunneling spectroscopy, the triple-points only provide small keyholes through which one can glimpse the full spectrum. By instead parallel-coupling the DQD to source and drain, it becomes possible to track states also far away from these points, and to decouple level detuning from the source-drain bias. Such a modification considerably expands the spectroscopic information that can be gained, and is the basis for the work presented here.

The general approach we adopt in this study follows the pioneering works by Hatano *et al.*, who used hybrid vertical-lateral GaAs DQDs, parallel-coupled with hard-wall barriers to source and drain, and with an inter-dot tunnel coupling tunable with side-gates [13-15]. There, the authors studied the evolution of various states as function of electron numbers, level detuning and inter-dot tunnel coupling. However, the spectroscopic resolution and tuning range of spin states was limited due to a relatively weak quantum confinement and small |*g*|-factor of GaAs.

In this work, we instead investigate parallel DQDs formed in nanowires of InAs [16], a material that provides considerably stronger quantum confinement, and larger |*g*|-factors, allowing better-resolved excited states and much wider range of spin tuning [17]. With electron numbers starting at zero, we map out excited states up to the first spin-paired shell in each QD as a function of level detuning. Focusing on the 1–2 electron transition near the (1,0)-(1,1)-(2,0) triple-point, we follow the evolution of the (1,1) triplet (T) state, and the hybridization of the (1,1) and (2,0) singlet (S) states. With an external magnetic field, and a system tuned to weak singlet hybridization, transport through all spin-split T(1,1) states is clearly resolved. Here, a S-T mixing of the S(2,0) and T(1,1) through SO-interaction allows violation of the normal spin selection rules as these states come close in energy. In particular we find that for a given inter-dot tunnel coupling, the energy for the S(2,0)-T(1,1) avoided crossing is considerably higher compared to the S(1,1)-T(1,1) avoided crossing in the same system [17]; where the former is a result of a first-order tunneling process between the QDs, and the latter is second-order.

Transport calculations based on a DQD model with one orbital in each QD, and with parameters extracted from the experiment, fully reproduce the measured transport spectrum including the ground and excited state evolution as a function of detuning and external magnetic field.

Finally, we also investigate the 2–3 electron transition, and observe a mirroring of the transition energies compared with the 1-2 electron transitions. This is expected as these transitions involve transport from 2-electron to 3-electron states, where the latter can be viewed as hole equivalents of the 1-electron doublet (D) states. By resolving all relevant transitions involving the first pair of orbitals, we thus provide experimental verification of theoretical predictions found in the literature on the evolution and interaction of these states [18].

## II. RESULTS AND DISCUSSION

The QD structure studied in this work is obtained by controlling the crystal structure of InAs during epitaxial growth of nanowires [16, 19, 20]. The nanowire leads, and the QD itself, have zinc-blende (ZB) crystal-phase, whereas the tunnel barriers consist of 20-30 nm long segments of wurtzite (WZ) InAs [17]. The enclosed ZB segment has a hexagonal shape, with an axial length of around 5 nm and a diameter of 70 nm, Fig. 1(a).

The WZ tunnel barriers provide reasonably hard-wall barriers to source and drain, with a conduction band offset of approximately 100 meV [19-21]. A set of three gate electrodes – a back-gate (BG) and two side gates (L, R) – are used to control the potential profile within the ZB segment, Fig. 1(b). The potential can be modulated such that the segment can host either one or two QDs, with a one-electron inter-dot tunnel coupling, $t$, tunable over an order of magnitude, 0.15 meV $< t <$ 1.5 meV [16,17]. We focus this study on a regime with rather weak inter-dot tunnel coupling, 0.15 meV $< t <$ 0.30 meV and on transport through the *very first* electron orbital in each QD, Fig. 1(c). In contrast to most other works, we thus do not rely on the assumption that lower-energy spin-paired electrons can be neglected. In serial

DQDs, where current levels are much lower, such characterization typically requires the presence of an integrated charge detector [22]. Moreover, owing to a relatively high source-drain resistance of the device, $R > 500$ k$\Omega$, and a low electron temperature, $T_{el} < 100$ mK, both the lifetime- and thermal-broadening of the levels are much smaller than the single-particle level spacing, which allow high-resolution spectroscopy.

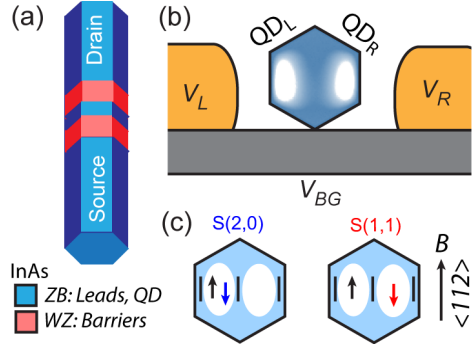

FIG. 1. (a) Schematic illustration of the InAs nanowire QD structure, with zinc-blende (ZB) crystal phase in the QD and leads, and wurtzite (WZ) tunnel barriers. (b) Voltages applied to the side-gates ($V_L$, $V_R$) and back-gate ($V_{BG}$) control the internal electrostatic potential and split the QD into two coupled electron pockets (white). (c) Illustration of two different two-electron states of the DQD: singlet S(2,0) and singlet S(1,1), where the small arrows denote spin. The large arrow indicates the approximate $B$-field orientation relative to the ZB crystal in the measurements.

Figure 2a shows a charge stability diagram ($dI/dV_L$) in the weak inter-dot tunnel coupling regime ($t = 0.15$ meV), at $V_{SD} = -1.9$ mV for a first cool-down (A). The bias voltage is here chosen such that $eV_{SD} < 2t+U_{12}$, where $U_{12}$ is the inter-dot Coulomb energy. We note that a higher $V_{SD}$ would result in overlapping conductance stripes from different charge states near the triple points, and complicate the analysis. The numbers given within the stretched-out honeycombs indicate the electron populations in the left and right QDs, controlled by the side gate voltages, $V_L$ and $V_R$. Corresponding higher resolution plots are provided in Figs. 2(b-e) for all four triple-point pairs.

When the energy of the lowest orbitals of the two QDs align in Fig. 2(d), we observe an avoided crossing corresponding to a bonding and an anti-bonding state resulting from hybridization of the D(1,0) and D(0,1) states. The same behavior is observed in Fig. 2(c) when the energy corresponding to a doubly occupied lowest orbital in one QD aligns with the energy of a singly occupied lowest orbital in the other. This results in hybridization of D(2,1) and D(1,2) states, which also can be seen as the bonding and anti-bonding states of a single hole in the DQD.

In Figs. 2(b,e) we observe similar avoided crossings, but here with an additional line cutting through them. In the lower left of Fig. 2(b), this straight line represents transport from D(1,0) to T(1,1), whereas the avoided crossing is now a result of transport from D(1,0) to hybridized S(2,0) – S(1,1) states. We note that the avoided singlet crossings and triplet lines are not equally visible in Fig. 2. Asymmetries in the tunnel barrier resistances to source and drain affect the overall conductance of each orbital (cf. Fig. 5 in the Appendix), and also whether excited states are more visible at positive or negative bias.

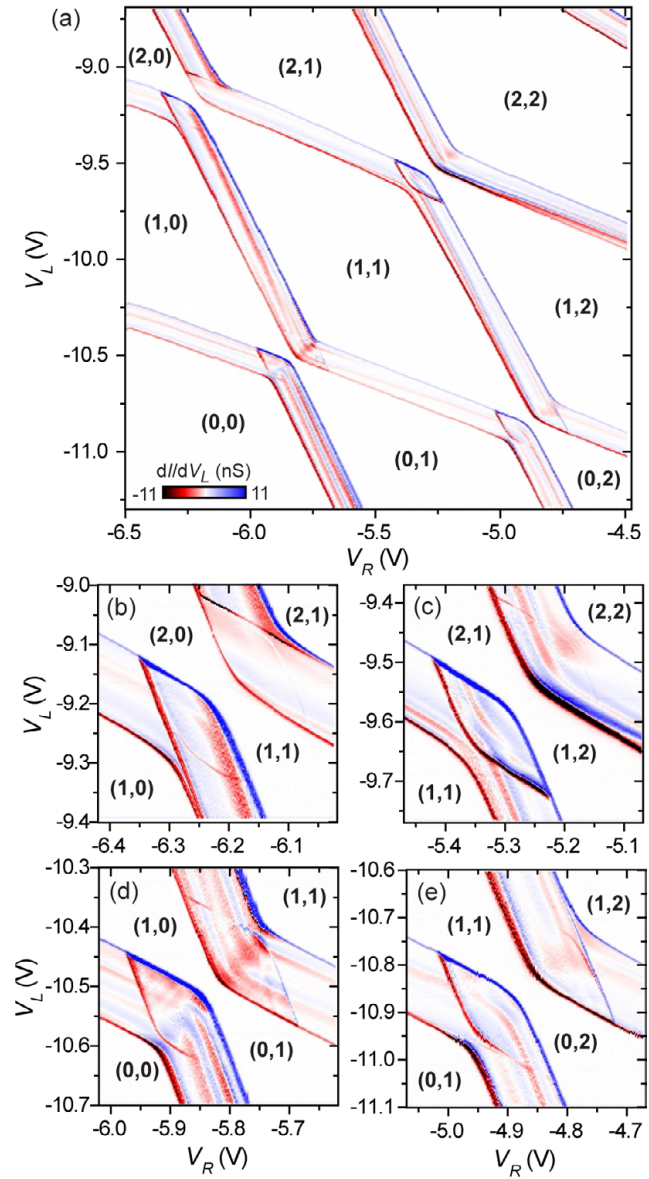

FIG. 2. Data from cool-down A: $dI/dV_L$ vs. $V_L$ and $V_R$ for $V_{BG}$ = +0.5 V, $V_{SD}$ = -1.9 mV, and $B$ = 0 T. (a) Overview diagram of the crossings of the first spin-degenerate orbital in each QD, where the sharp features appearing inside the conductance stripes originate from transport through excited states. (b)-(e) Higher resolution measurements of the level-crossings in (a), where anti-bonding and spin triplet excited states are resolved.

Next we focus on the (1,0)-(1,1)-(2,0) triple-point, and investigate how energy levels are affected by detuning and an external magnetic field, $B$. The same DQD is studied, but now in a different cool-down (B). Figure 3(a) provides an overview of the involved states for the 1–2 electron transition, and for the lowest orbital in each dot. The star (*) symbol indicates transitions that start in the D(1,0) spin-down state, [D(↓,0)] which becomes an excited state for $B > 0$.

Figures 3(b, c) show how the energies of the relevant 1- and 2-electron states change with one-electron detuning, $\Delta = E(0,1) - E(1,0)$, using a DQD model with one spin-degenerate orbital in each QD [18] and parameters extracted from the experiments. We note that asymmetric charge states, D(1,0) and S(2,0), are sloped, whereas symmetric (1,1) states are straight with respect to detuning. An external magnetic field Zeeman splits the D(1,0) states in (b), and also the T(1,1) states in (c). In order to better visualize the states and their mixing, pure S(1,1), T(1,1), and S(2,0) states are indicated with red, green, and blue color respectively. A mixed color represents the degree of hybridization of states, which is prominent near the avoided crossings in Fig. 3(c). Here, S(2,0) and S(1,1) anticross with $\Delta E = 2\sqrt{2}t$, whereas T(1,1) and S(2,0) anticross as a result of spin-orbit coupling, with an energy $2\Delta_{ST}$.

Figure 3(d) shows the results of a transport calculation for the 1-2 electron transition at $B$ = 2 T as a function of level detuning, $\Delta + \delta$, with $\delta$ = 10.7 meV such that $\Delta + \delta = 0$ at the center of the singlet-singlet crossing. We have used the eigenstates of the DQD model as input to a master equation based on a leading order perturbation expansion in the tunnel couplings between the QDs and the source and drain leads. The master equation is solved for the non-equilibrium populations, from which the current is then calculated. A description of the model can be found in the Supplemental Material of Ref. 17 and the input parameters used are provided in Table 1. We note that there is a rotation of the features compared to the 2-electron states in Fig. 3(c), which is a consequence of detuning of the 1-electron states. Furthermore, some additional lines appear because of transport processes starting from spin-split 1-electron excited states (discussed below).

**TABLE 1**

Numerical values used in the transport model. The parameters $t$, $V_x$ (exchange energy), $U_{11}$, $U_{22}$ (intradot Coulomb energies for QDs 1 and 2), $U_{12}$ (interdot Coulomb energy), $\alpha$ and $|g|$ are input parameters to the QD Hamiltonian while $T$ (temperature), $t_1$, $t_2$, $t_s$, and $t_d$ (coupling to source and drain respectively) are input parameters to the transport calculations. The Hamiltonian and all parameter definitions can be found in the Appendix.

| $t$ / meV | $V_x$ / meV | $U_{11}$ / meV | $U_{22}$ / meV | $U_{12}$ / meV | $\alpha$ | $|g|$ | $T$ / mK | $t_2 / t_1$ | $t_s / t_d$ |
|---|---|---|---|---|---|---|---|---|---|
| 0.14 | 0 | 14.8 | 13.3 | 2.58 | 0.6 | 8.9 | 46 | 1 | 2 |

Figures 3(e-g) shows the experimental $dI/dV_R$ plotted vs. energy and detuning for a 1–2 electron transition at $B = 0$, 1, and 2 T. Here, the $V_L$ and $V_R$ scales are locally converted to energy by extracting gate lever arms from Coulomb charge stability diagrams ($V_{SD}$ vs. $V_R$ and $V_L$), and the slopes of the (1,1) honeycomb borders, but also from the width of the conductance stripes set by $eV_{SD} = 2.5$ meV. In all of the figures, $E$ and $(\Delta + \delta)$ are set to zero in the middle of the avoided singlet crossing. More information about the extraction of gate lever arms is shown in Fig. 6, in the Appendix, where also Figs. 7, 8 show data before the conversion.

In the meeting point of the two conductance stripes in Fig. 3(e) we note a clear, avoided S(2,0)-S(1,1) crossing, and a triplet T(1,1) line that runs through it. From the vertical axis we can extract an energy for the avoided singlet crossing of 490 µeV, which corresponds to an inter-dot tunnel coupling $t = 170$ µeV. The scale in the energy conversion is corroborated by a Coulomb stability diagram ($V_{SD}$ vs. $V_{L,R}$) measured approximately through the triple points, shown in Fig. 7 in the Appendix, from which the T(1,1) and singlet anti-bond excited state energies can be extracted.

When a magnetic field is applied [here close to a <112>-type direction, Fig. 1(c)], the T(1,1) states Zeeman split with $\Delta E_Z = g\mu_B B$, where $\mu_B$ is the Bohr magneton and $g$ the effective $g$-factor. An average $|g|$-factor of 8.9 is extracted from the spin-splitting of the T(1,1) transitions at $B$ = 1 and 2 T, consistent with earlier observations [5, 17]. Three avoided crossings are visible in Figs. 3(f, g) in agreement with the predictions for a SO-coupling between S(2,0) and T(1,1) states. Transport from D(↑,0) to T-(1,1), which normally should be forbidden, is here possible close to the T-(1,1)-S(2,0) avoided crossing where these state are strongly hybridized and normal spin-selection rules are broken by the SO-coupling. The energy gap at the avoided S-T crossing ($2\Delta_{ST}$) is associated with the amplitude for electrons tunneling between the two QDs and flipping their spin. The magnitude of $\Delta_{ST}$ is thus expected to depend on the tunnel coupling $t$ between the two dots, as well as on the spin-orbit coupling which causes spin flips [17, 23]. In this study, because of the relatively small $t$ = 170 μeV, we extract correspondingly small values for the two avoided S-T crossing energies, $\Delta_{ST}$ = 80 μeV.

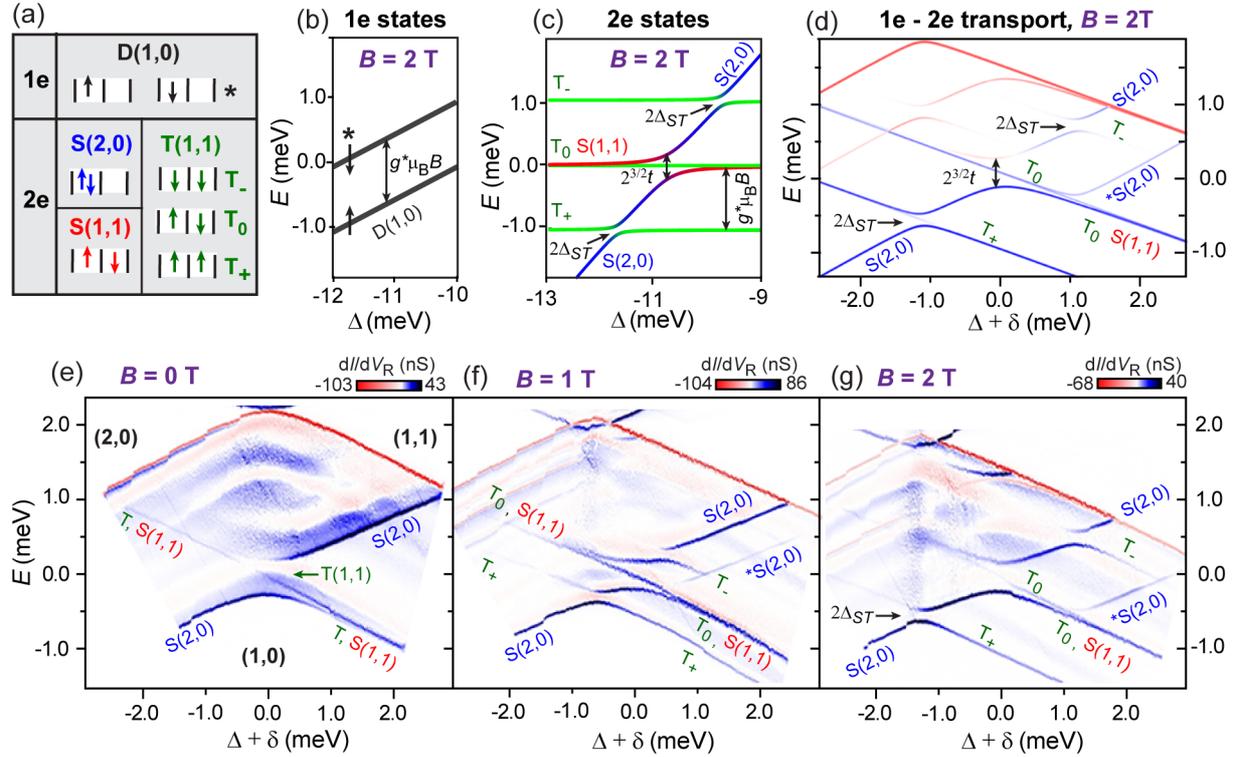

FIG. 3. Simulations and experimental data from cool-down B at $V_{SD}$ = 2.5 mV. (a) Simplified representation of the spin configurations for the relevant 1- and 2-electron states. (b, c) Calculated 1-electron (b) and 2-electron (c) energies vs. detuning near the (1,0)-(1,1)-(2,0) triple-point. (d) Calculated $(dI/dV_L + dI/dV_R)/2$ at $V_{SD}$ = 2.5 mV and $B$ = 2 T, near the (1,0)-(1,1)-(2,0) triple-point, with detuning shifted by δ = 10.7 meV. (e-g) Experimental $dI/dV_R$ plotted against detuning (set to 0 at the S-S crossing) and energy for $V_{SD}$ = 2.5 mV, $V_{BG}$ = -0.5 V and different applied $B$-fields. (e) S(2,0)-S(1,1) avoided crossing, with T(1,1) running diagonally through the hybridization gap at $B$ = 0 T. (f, g) New lines appear due to Zeeman splitting of D(1,0) and T(1,1) states at $B$ > 0. The line marked T-

requires a spin-flip and is visible only close to the T₋-S(2,0) avoided crossing. The weak S(2,0) replica, marked *S(2,0), involves transport processes starting from the excited D(↓,0) state. As noted in the main text, there are other transitions involving the D(↓,0) state that have the same energies as the lines indicated $T_+$ and $T_0$ in (f, g).

It is important to note that, in addition to the ground-state (GS) to excited-state (ES) transitions marked in Fig. 3(f, g), there are a few transitions which become possible at the same energy but that start in the spin-down ES, denoted D(↓,0) [2]. Here, the D(↓,0)-T₋(1,1) ES-ES transition becomes possible at the same energy as the D(↑,0)-$T_0$(1,1) GS-ES transition. Similarly, the D(↑,0)-T₊(1,1) GS-GS transition (denoted $T_+$) becomes possible at the same energy as both the D(↓,0)-S(1,1) and the D(↓,0)-$T_0$(1,1) ES-ES transitions. However, usually the GS is occupied with much larger probability than the ES, and we therefore expect that the main contribution to the transport comes from the transitions that start from a GS. From transport theory in Fig. 3(d) it is clear that the latter ES-ES transitions cut through the lower avoided S-T₊ crossing energies. As the lines are not resolved in the experimental data, we can conclude that their contribution is small compared to GS-GS transport. The calculations do not include relaxation of an ES due to other effects than electron tunneling, which can explain why ES-ES transitions are somewhat stronger than in the experimental data.

For the case of transitions involving S(2,0), an even more direct comparison of the tunneling probabilities from a GS doublet and an ES doublet is possible. In Figs. 3(f, g) there is a weaker line running parallel to the S(2,0) transition, marked *S(2,0), with almost identical spin-splitting energy as the T(1,1) transitions, but with a conductance $G_{*S(2,0)} \sim 0.3 G_{S(2,0)}$. This weaker line represents tunneling from the spin-down D(↓,0) ES to the S(2,0) ES. In Fig. 8 in the Appendix we provide multiple Coulomb stability diagrams obtained at line cuts across various features in Fig. 3(f) at $B = 1$ T, such as the spin-split D(1,0)-S(2,0) transition.

In the last data set we focus on the 2–3 electron transition, which is better resolved at the (1,1)-(0,2)-(1,2) triple-point, and at negative $V_{SD} = -2.5$ mV (Fig. 4). A general conclusion is that we observe a mirroring of the spectrum relative to Fig. 3, which is expected due to the two-fold spin degeneracy of each involved orbital. Transitions from a three-electron state to a two-electron state by removing an electron can here be seen as the hole-equivalent of transitions from a one-electron state to a two-electron state by adding one electron. Figure 4(b), recorded at $B = 0$ T, reveals a very clear avoided singlet-crossing, and a T(1,1) state that cuts through the gap. However, as the inter-dot tunnel coupling now is stronger ($t = 300$ μeV), the S-T avoided crossings at $B = 1$ T partly overlap with the avoided singlet-crossing, and are difficult to resolve. At a higher field, $B = 2$ T, all three avoided crossings can be resolved, and we find that the spectrum is similar to Fig. 3(f), although mirrored in both planes. For this data set we extract a $\Delta_{ST} = 130$ μeV, and an average |g|-factor of 9.1 for the T(1,1) transitions.

Most transitions in Fig. 4(d) connect a 2-electron state and the 3-electron GS, which is D(↑,2) for $B > 0$. However, similar to Fig. 3, there are a few exceptions, such as the line marked *S(0,2) which now ends in the D(↓,2) ES, a transition which here is at higher energy than the S(0,2)-D(↑,2). Similarly, we note a reverse order of the triplet transition energies, with T$_+$(1,1) being the highest, explained by that T$_+$(1,1) is the 2-electron GS.

From the spin-splitting of the S(2,0) and S(0,2) transitions in Figs. 3(e, f) and 4(c, d), we can directly access values for the |g|-factors of the first orbital in each QD, D(1,0) and D(0,1). We find that they are rather similar; $|g_L| = 9.2$ and $|g_R| = 9.1$, with values that are in agreement with those observed for the T(1,1)-splitting. However, as they are extracted at different points in the honeycomb diagram, for different $V_L$ and $V_R$, the values only provide an indication to the magnetic field response of the two QDs.

Finally we note that for a given $\Delta_{ST}$ the associated $t$ is here a factor of 5-10 smaller than in Nilsson *et al*. [17], where the S(1,1) – T(1,1) avoided crossing was studied for the same DQD. In that case the two states mix through second order tunneling processes between the QDs, scaling with $\Delta_{ST} \sim tt_{so}/U$, where $t_{so} = \alpha t$ is the tunneling element between the dots involving a spin-flip and $U$ is a function of the inter- and intradot Coulomb energies; whereas S(2,0) – T(1,1) mixes in first order ($\alpha t$). Additionally we find a significant difference in the spin-orbit coupling term, $\alpha$, in the model for the two cases, which could be due to a different spin-orbit vector induced by changes in the electric field and the shapes of the QDs.

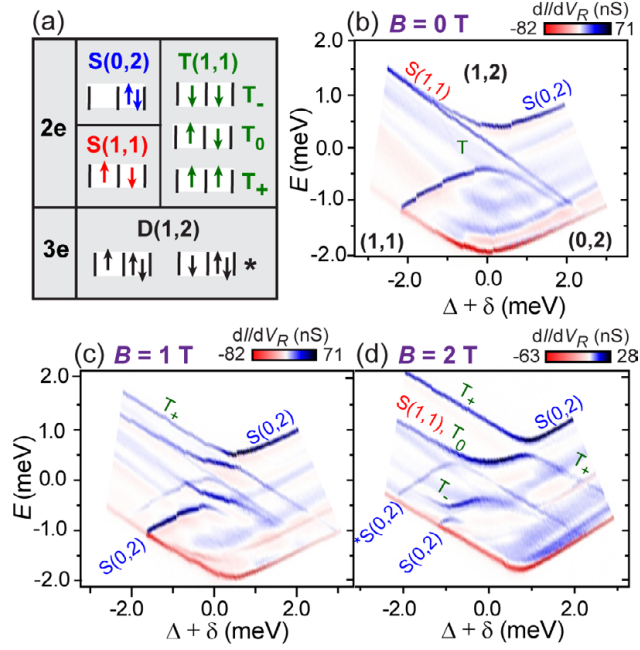

FIG. 4. Data from cool-down B. (a) Spin configurations for the relevant 2- and 3-electron states. (b) $dI/dV_R$ for the 2-3 electron transition at the (1,1)-(0,2)-(1,2) triple point, for $V_{BG}$ = -0.5 V, $V_{SD}$ = -2.5 mV, and $B$ = 0 T. (c-d) For $B > 0$, the T(1,1) states spin split. A mirrored symmetry is found between the 2-3 and 1-2 electron transitions [cf. Fig. 3(g)], which is expected due to the two-fold spin degeneracy of the two interacting orbitals.

### III. SUMMARY AND CONCLUSIONS

In summary, we provide high-resolution excited state spectroscopy of few-electron transitions in parallel InAs DQDs. Very strong and stable QD confining potentials allow transport studies with significant QD level detuning, starting from completely empty QDs. An external magnetic field spin-splits the T(1,1) states, and we can resolve transport through $T_+$, $T_0$, as well as the $T_-$ state for energies where $T_-$ is hybridized with S(2,0) via spin-orbit coupling. Our results are reproduced by a master-equation calculation assuming two parallel single-level QDs. For the same DQD, we find a considerably larger SO-induced avoided crossing for S(2,0)-T(1,1) compared to S(1,1)-T(1,1), explained by that the latter is coupled by a second-order tunneling process between the QDs. A more exact picture of the SO-coupling, would be obtained by studying the B-field angular dependence of the S-T avoided crossing energy, similar to Nadj-Perge *et al.* [24].

**ACKNOWLEDGEMENTS**

This work was carried out with financial support from NanoLund, the Swedish Research Council (VR), the Swedish Foundation for Strategic Research (SSF), the Crafoord Foundation, and the Knut and Alice Wallenberg Foundation (KAW).

**APPENDIX**

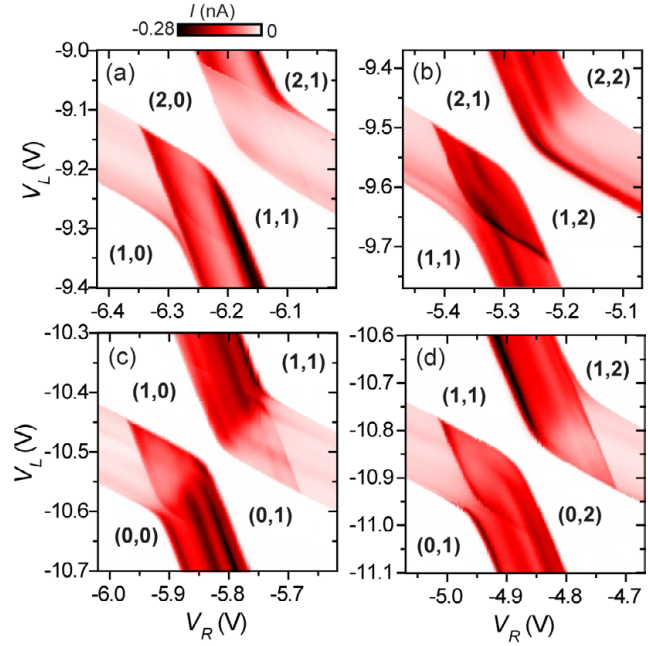

FIG. 5. (a-d) Current as function of side-gate voltages at $V_{SD}$ = - 1.9 mV for cool-down A, corresponding to Figs. 2(b-e). $QD_R$ has higher conductance than $QD_L$. Therefore, in (a), if S(2,0) is the ground state (both electrons in $QD_L$), then the 2-electron excited states that involve $QD_R$ [(1,1)-states] provide a significant conductance increase. However, if S(1,1) or T(1,1) is the ground state, then the S(2,0) excited state only gives a small contribution.

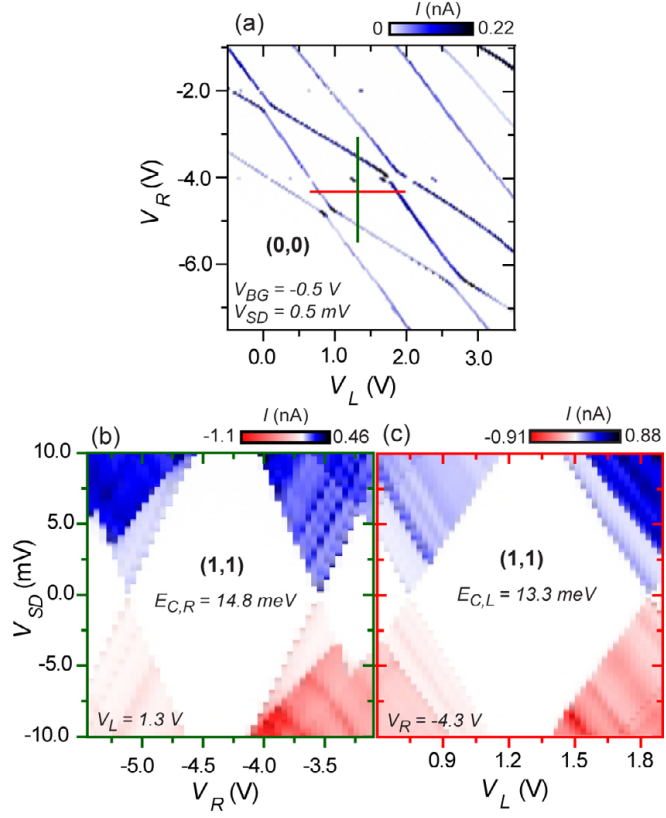

FIG. 6. (a) Honeycomb stability diagram at $V_{SD} = 0.5$ mV, $V_{BG} = -0.5$ V and $B = 0$ T for cooldown B. (b, c) Coulomb stability diagrams, $dI/dV_{SD}$, recorded along $V_R$ and $V_L$, which provide the charging energies and lever-arms of the two QDs. The cross-capacitance lever-arms ($V_{gate-QD}$) are extracted from the slopes of the honeycomb borders in (a), and are $V_{L-L} = 9.9$, $V_{L-R} = 5.2$, $V_{R-R} = 12.2$, $V_{R-L} = 9.2$ meV/V.

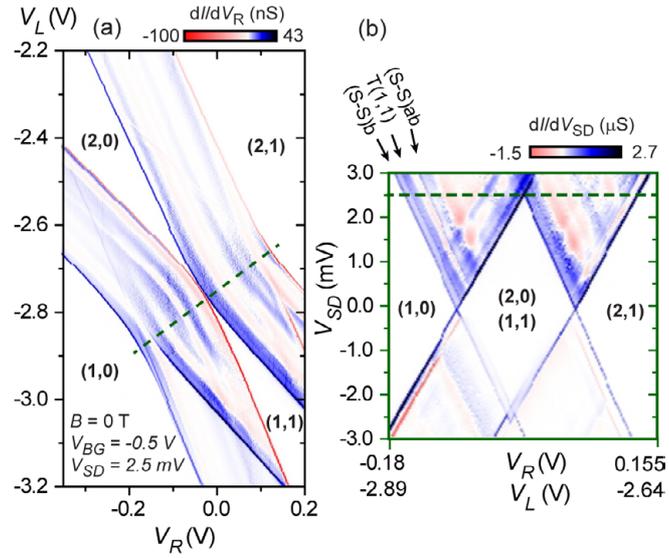

FIG. 7. (a) d$I$/d$V_R$ as function of side-gate voltages for cool-down B near the (2,0)-(1,1) degeneracy point. (b) Stability diagram (d$I$/d$V_{SD}$) recorded through the two triple-points along the side-gate vector indicated in (a), providing information on how ground and excited states shift with $V_{SD}$. The subscript "b" denotes bonding orbital, whereas, "ab" denotes anti-bond.

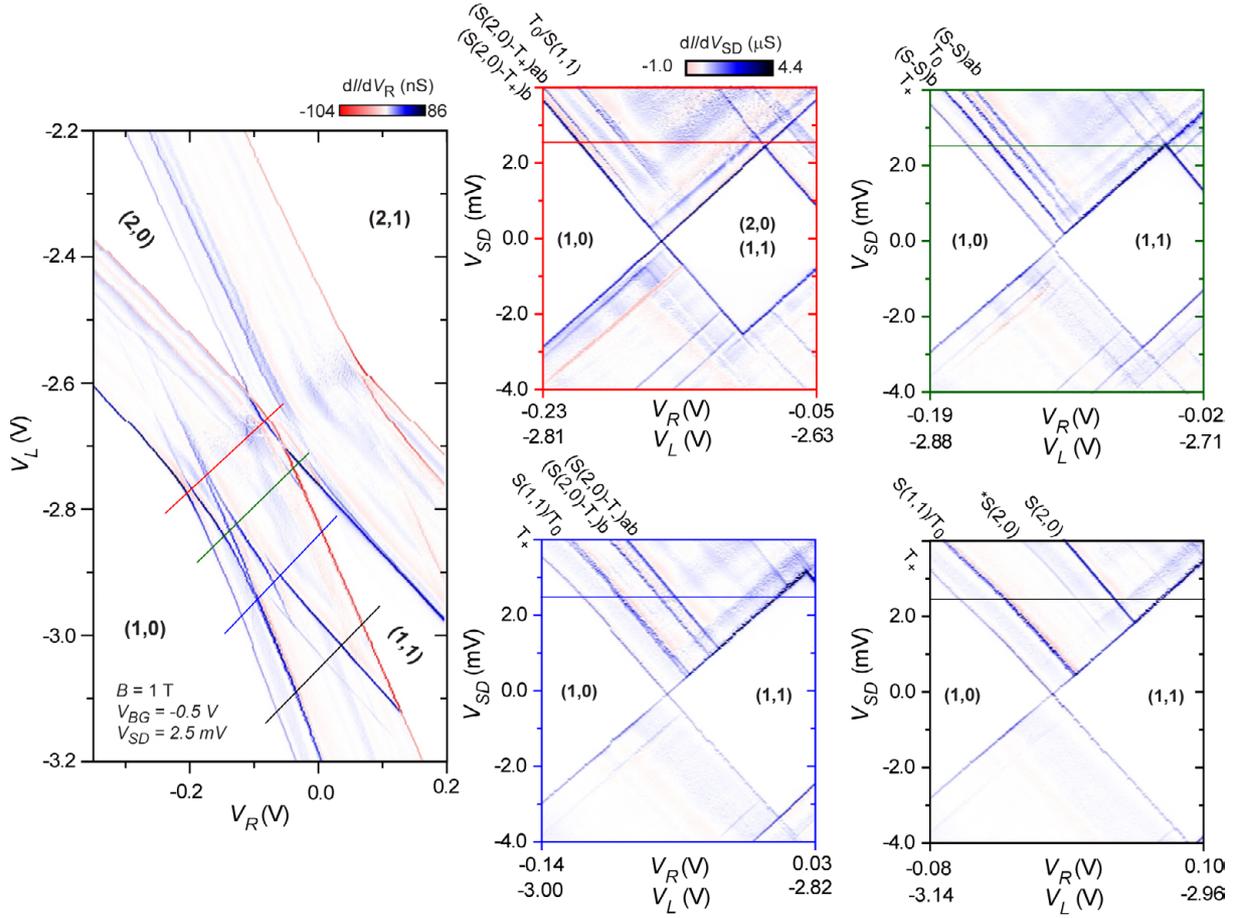

FIG. 8. Plot of d$I$/d$V_R$ as function of side-gate voltages for cool-down B near the (2,0)-(1,1) degeneracy point at $B = 1$ T. The colored lines indicate vectors along which stability diagrams (d$I$/d$V_{SD}$) are recorded. These provide information on how ground and excited states shift with $V_{SD}$. The subscript "b" denotes bonding orbital, whereas, "ab" denotes anti-bond.

## IV.  MODEL

The Hamiltonian for the QDs used in the calculations is given by

$$H = H_0 + H_{SO} + H_B$$

with

$$H_0 = \sum_{\substack{i=1,2 \\ \sigma=\uparrow,\downarrow}} \epsilon_i c^\dagger_{i\sigma} c_{i\sigma} - \sum_{\substack{i\neq j \\ \sigma=\uparrow,\downarrow}} t c^\dagger_{i\sigma} c_{j\sigma} + \sum_{\substack{i=1,2 \\ i\neq j \\ \sigma=\uparrow,\downarrow}} \left[ \frac{U_{ii}}{2} \delta_{\sigma\neq\sigma'} c^\dagger_{i\sigma} c^\dagger_{i\sigma'} c_{i\sigma'} c_{i\sigma} + \frac{U_{ij}}{2} c^\dagger_{i\sigma} c^\dagger_{j\sigma'} c_{j\sigma'} c_{i\sigma} - \frac{V_x}{2} c^\dagger_{i\sigma} c^\dagger_{j\sigma'} c_{j\sigma} c_{i\sigma'} \right],$$

$$H_{SO} = \sum_{\substack{i,j=1,2 \\ i \neq j}} (-1)^i it_{SO}(-c_{i\uparrow}^\dagger c_{j\uparrow} + c_{i\downarrow}^\dagger c_{j\downarrow})$$

and

$$H_B = \sum_{i=1,2} \frac{|g|\mu_B B}{2}(c_{i\uparrow}^\dagger c_{i\downarrow} + c_{i\downarrow}^\dagger c_{i\uparrow}),$$

where $\varepsilon_i$ is the one-electron energy on dot, $t$ the interdot tunnel coupling, $U_{ij}$ the charging energy, $V_x$ the exchange integral, $t_{so} = \alpha t$ the spin-orbit coupling parameter, $|g|$ the effective g-factor, $\mu_b$ the Bohr magneton and $B$ the external magnetic field. In addition, the temperature $T$, the amplitudes $t_1$ and $t_2$ for tunneling into dot 1 and 2, and the amplitudes $t_s$ and $t_d$ for tunneling to/from source and drain are used as input parameters to the transport calculations. A full description of the transport model can be found in the Supplemental material to Ref. 17.